\documentclass[A4, 12pt ] {article}
\usepackage[english]{babel}
\begin{document}

\title{About MathPartner web service \thanks{ This paper was published in:   Gennadi Malaschonok and Ivan Borisov. About MathPartner web service. International conference Polynomial
Computer Algebra. St.Petersburg, PDMI RAS, 2014. P. 50-54.} }
\author{Gennadi Malaschonok and Ivan Borisov}
\date{}

\maketitle

\begin{abstract}
The report is devoted to the current state of the MathPartner computer algebra web project.
We discuss the main directions of development of the project and give several examples of using it to solve selected problems.
\end{abstract}

MathPartner is a web service for symbolic and numerical computations.
Mathpar language is a \TeX{}-like language also called \textrm{A}\TeX{} (stands for Active \TeX{}).

Service provides the user with a workbook in which you can type the text and use all mathematical services.
User can perform numerical and symbolic calculation; solve differential equalities; 
do matrix and polynomial computation; plot graphics of 2D and 3D functions, interact with 3D graphics, 
make animation and save graphics for further usage.

One of the very important features of this mathmatical servise is the explicit declaration of mathematical environment called Space. All calculations are made at the specified Space and user can change it on demand.

\textbf{Space environment}

\texttt{SPACE} is defined by numeric sets, algebraic operations in these sets and variable names which appear in polynomials and functions. By default, a Space of the four real variables R64[x, y, z, t] is defined. This is a polynomial ring with coefficients in the field of real numbers with double precision, variables have the ordering: $x < y < z < t$. User can change the space and ordering of variables. For example, \texttt{SPACE = Q[y,x]} is a ring of two variables $x$ and $y$, $y< x$, over rational numbers.

\textbf{Mathpar language basics}

All functions, operators and reserved words start with backslash (\textbackslash) symbol, e.~g.: \texttt{\textbackslash sin(x)}, \texttt{\textbackslash infty}. Most of syntax conventions in Mathpar language came from \TeX{} math mode: function names, upper and lower indices, Greek letters, math symbols etc. Names of elements of noncommutative algebras also have to start with backslash and the first letter capitalized e.~g.: \texttt{\textbackslash A}.

Expressions are separated with a semicolon (;) or text comments, which are enclosed in quotation marks ("). To see the results use the \texttt{\textbackslash print(exp1, ..., expN)} command with the names of expressions as arguments. If the list of statements does not contain a print-like operator (\texttt{print} itself, \texttt{prints}, \texttt{plot}, etc.) the result of the last statement will be shown.

All math text that is in the workbook, can be divided into parts and is located in a separate window. Each part is compiled and executed independently.

\textbf{User interface}

At the top navigation bar there are links to main pages of Mathpar: welcome page, workbook, help, about. Also you can change language  (currently English and Russian are available) and download handbook which contains help topics in PDF format with user guide  for those who use Mathpar in their study.

The workbook  has left sidebar with button groups located on collapsible panels. These buttons help you to insert commands so you don't have to search for a syntax of needed command all the time. There are here also indicators of current free server memory and working SPACE, some utility panels for uploading files with user data and for downloading results in the PDF format, user authentication etc.
 
 Each windows, which contains one section of math text, has a toolbar at the top with buttons from left to right: run section; toggle output  between Mathpar mode  and \LaTeX{} mode; add a new section below; clear all user-defined expressions; remove section. Mathpar mode lets you to edit input text or copy result for further usage. \LaTeX{} mode gives you the opportunity to see the beautiful mathematical formulas.

\textbf{Functions} 

To calculate the value of a function at a point use the function: \\ \texttt{\textbackslash value(f, [var1, var2, ..., varn])}, \\  where \texttt{f} is a function, and \texttt{var1,var2, ...,varn} -- values of the space variables.
Example for obtaining the value of expression  
$g= \mathbf{sin}(1^2 + \mathbf{tg}(2^3 + 1))$:  
\begin{verbatim}
SPACE = R64[x, y];
f = \sin(x^2 + \tg(y^3 + x));
g = \value(f, [1, 2]);
\print(g);
\end{verbatim}
gives $g = 0.52$

You can also substitute variables \texttt{var1,var2, ...,varn} with more complex expressions than a numbers:
\begin{verbatim}
SPACE = R[x, y];
f = x^2 + y^2;
g = \value(f, [\sin(x), \cos(x)]);
\Factor(g);
\end{verbatim}
gives $1$

Here \texttt{\textbackslash Factor} is a function for simplification of trigonometric and logarithmic functions and their compositions.

Example of function integration and differentiation:
\begin{verbatim}
SPACE = Q[x];
f = (2x^2 + 1)^3;
l = \int(f) d x;
dl = \D(l, x);
d2l = \D(l, x^2);
\print(f, l, dl, d2l);
\end{verbatim}
which gives:
$f = 8x^6+12x^4+6x^2+1;\\
l = (8/7)x^7+(12/5)x^5+2x^3+x;\\
dl = 8x^6+12x^4+6x^2+1;\\
d2l = 48x^5+48x^3+12x;$

You can notice that the designation \texttt{\textbackslash D(f, var\^{ }n)} is used to find an n-th order derivative of function \textbf{f} by variable \textbf{var}.

\textbf{Polynomials}

Calculation the value of polynomial at the point is the same as for function:
\begin{verbatim}
SPACE = R[x, y];
f = x^2 + 5x(y^3 + x);
g = \value(f, [1, 2]);
\end{verbatim}
$g=46.00$

Solution of algebraic equation:
\begin{verbatim}
SPACE = C64[x];
FLOATPOS = 2;
b = \solve(x^4 + 2x + 1 = 0);
\end{verbatim}
$[(0.77+1.12 \backslash i),(0.77-1.12 \backslash i),-0.54,-1.0]$

Solution of algebraic inequality:
\begin{verbatim}
SPACE = R[x];
b = \solve((x + 1)^2(x - 3)(x + 5) \ge 0);
\end{verbatim}
$(-\infty, -5]\cup\{-1\}\cup[3, \infty);$

Groebner basis of polynomial ideal:
\begin{verbatim}
SPACE = Z[x, y, z];
\gbasis(x^4y^3 + 2xy^2 + 3x + 1, x^3y^2 + x^2, x^4y + z^2+xy^4 + 3);
\end{verbatim}
$[z^2-x^4+3x^2-10x+9,y-9x^4-3x^3-x^2-81x+27,x^5+9x^2-6x+1];$

Solution of the polynomial system:
\begin{verbatim}
SPACE = R[x, y];
\solveNAE(x^2 + y^2 - 4, y - x^2);
\end{verbatim}
The result will be in form of matrix with columns correspond to variables, and each row is a solution.

$\left(\begin{array}{cc}1.60\mathbf{i} & -2.56 \\ 1.24 & 1.56 \\ -1.60\mathbf{i} & -2.56 \\ -1.24 & 1.56 \end{array}\right)$

\textbf{Tropical mathematics}

There are several tropical semifields and semirings in MathPartner such as: \texttt{ZMaxPlus}, \texttt{RMinMult}, \texttt{R64MinMax}. Check the Help for a list of all available tropical algebras.

Name of tropical algebra consists of three parts: a numerical set; an operation corresponding to plus; an operation corresponding to times.

Each algebra has elements \textbf{0} and \textbf{1}, some om it has elements $-\infty$ and $\infty$.

Example with simple operations:
\begin{verbatim}
SPACE = ZMaxMult[x, y];
a = 2; b = 9;
c = a + b; d = a b;
\print(c, d)
\end{verbatim}
$c=9;\\
d=18;$

Let $A$ -- matrix, $x$ -- column vector of unknown variables, $b$ -- column vector of constant terms.
To solve problems $A x = b$, $A x \le b$, $A x = x$, $A x \oplus b = x$ we have commands \texttt{\textbackslash solveLAETropic(A, b)}, \texttt{\textbackslash solveLAITropic(A, b)}, \linebreak \texttt{\textbackslash BellmanEquation(A)}, \texttt{\textbackslash BellmanEquation(A, b)} respectively.

Two more commands are implemented for graphs theory. The first one is for searching least distances between all nodes of the graph \texttt{\textbackslash searchLeastDistances(A)} where A is a matrix of distances between nodes. The second is \texttt{\textbackslash findTheShortestPath(A, i, j)} for constructing the shortest path between i-th and j-th nodes.

For additional information about MathPartner service see [2-4].

\end{document}